\documentclass[aps,twocolumn,prl,showpacs,color,psfig,epsf]{revtex4}
\usepackage{amsmath}
\usepackage{color}
\usepackage{amsfonts}
\usepackage{epsf}
\usepackage{graphicx}
\usepackage{epstopdf}
\usepackage{ulem}

\begin{document}

\title{Hydrodynamics defines the stable swimming direction of spherical squirmers in a nematic liquid crystal}

\author{J. S. Lintuvuori$^1$, A. W{\"u}rger$^1$ and K. Stratford$^2$}
\affiliation{$^1$Laboratoire Ondes et Mati{\`e}re d'Aquitaine, Universit{\'e} de Bordeaux \& CNRS, 33405 Talence, France \\
$^2$ EPCC, University of Edinburgh, UK
}

\begin{abstract}
 We present a study of the hydrodynamics of an active particle---a model squirmer---in an environment with a broken rotational symmetry: a nematic liquid crystal. By combining simulations with analytic calculations, we show that the hydrodynamic coupling between the squirmer flow field and liquid crystalline director can lead to re-orientation of the swimmers. The preferred orientation depends on the exact details of the squirmer flow field. In a steady state, pushers are shown to swim parallel
  with the nematic director while pullers swim perpendicular to the nematic director. This behaviour arises solely from hydrodynamic coupling between the squirmer flow field and anisotropic viscosities of the host fluid. Our results suggest that an anisotropic swimming medium can be used to characterise and guide spherical microswimmers in the bulk.
  

\end{abstract}

\maketitle


Active materials use internal energy resources to propel themselves and have recently emerged as a topical research area within physics~\cite{Ramaswamy2010, Marchetti2013}. A natural example of an
active systems is provided by swimming bacteria, while artificial microswimmers can be realised by self-propelling Janus particles~\cite{brown14, brown15, ebbens12b, sabass10, ebbens12, ebbens14, wang15, das15}. One big challenge is to control and direct the swimmers at the microscale. Success
here could allow one to harness swimmers to do work, and it could lead to significant technological possibilities, for example, direct microengineering of new materials. 

Various possibilities have been explored in order to guide active particles. The most obvious one is to use confining walls, as both bacteria~\cite{Lord63, berke08} and artificial swimmers~\cite{ishimoto13, zottl14,li14, brown15, das15, lintuvuori16} are known to be attracted to surfaces, and swim near them. Motion along predefined pathways can be obtained by topographical patterns \cite{Sim16} or chemical functionalisation \cite{Usp16} of the surface. Force-free localization and steering of laser-powered Janus particles have been achieved by dynamical feedback \cite{Bre14} or by spatial modulation of the laser beam \cite{Loz16}, which exerts a torque on the moving particle \cite{Bic14}.

An alternative route to control the swimmers in the bulk is to use an anisotropic swimming media~\cite{Arratia16}, {\it e.g.} a liquid crystal~\cite{Lavrento16}. 
  Recent experiments of colloidal particles have demonstrated electrophoretic propulsion of spherical colloids in nematic LC~\cite{Lavrento16, Hernandez13}. Whereas rod-like bacteria
  are observed to swim along the direction set by the nematic director $\hat{\mathbf{n}}$ ~\cite{Smalyukh08, Kumar13, Aranson14}. Experimental applications include the self-assembly of bacteria dispersed in a nematic LC~\cite{Abbott14}, transport of colloidal cargo~\cite{Weibel15} and accumulation of the bacteria to topological patterns~\cite{lavrentovich1, lavrentovich2}. Theoretical predictions include anomalous diffusion~\cite{rik} and even backward swimming was predicted by theoretical calculations of Taylor-sheets in nematic LC~\cite{powers1,powers2}. 

In the case of rod-like swimmers ({\it e.g.} typical bacteria) the alignment is dominated by an elastic energy, which is minimised when the rods align their long-axis along $\hat{\mathbf{n}}$~\cite{Smalyukh08}, thus rod-like swimmers are always expected to swim following the nematic director. For isotropic swimmers ({\it e.g.} spherical bacteria or artificial Janus swimmers) this is not the case: in the limit of spherical particles the elastic torque vanishes.

In this letter, we study the dynamics of fully resolved {\it spherical} microswimmers in a nematic liquid crystal, by means of lattice Boltzmann simulations and analytical calculations, using a squirmer model~\cite{lighthill52}. Our simulations show that the steady state swimming direction depends of the nature of the swimming mechanism. Spherical {\it pushers} undergo stable swimming following the direction set by the nematic director. 
Strikingly, a {\it puller} swims in steady state in a direction orthogonal to the far-field $\hat{\mathbf{n}}$. 
Using analytical calculations we show that the reorientation is due to a hydrodynamic torque, arising from the coupling between the squirmer flow field and anisotropy of the liquid crystalline viscosities~\cite{miesowicz}. 
Further we show the reorientation rate scales linearly with the power of the squirmer flow field.
Our results provide a robust and easy way to manipulate self-propelling organisms directly at the microscale, allowing for example sorting of the swimmers based on their hydrodynamic nature. 


{\it Squirmer model:} To simulate the dynamics of an active particle in a liquid crystal we employ a lattice Boltzmann (LB) method~\cite{mikeLB}. We treat the self-propelling particle in the terms of a squirmer model~\cite{lighthill52}. The tangential (slip) velocity profile at the particle surface leading to the squirmer motions is given by~\cite{Magar03}
\begin{equation}
  u(\theta)= v_0 \sin (\theta)( 1+ \beta \cos \theta)
\label{eq:veltangential}
\end{equation}
where $v_0$ is a constant, $\beta$ the squirmer parameter, and $\theta$ the polar angle with respect to the particle axis ~\cite{ishimoto13}.

In the LB method a no-slip boundary condition at the fluid/solid interface can be achieved by using a standard method of bounce-back on links (BBL)~\cite{ladd1, ladd2}. When the boundary is moving ({\it e.g.} a colloidal particle) the BBL condition needs to be modified to take into account particle motion~\cite{ladd3}. These local rules can include additional terms, such as a surface slip velocity (Eq.~\ref{eq:veltangential}) leading to LB simulations of squirming motion~\cite{ignacio1,ignacio2}. 

{\it Liquid crystal model:} The nematic host fluid is described by a Landau -- de Gennes free-energy whose density can be expressed in terms of a symmetric and traceless order parameter tensor $\mathbf{Q}$ as ${\cal F} = F(Q_{\alpha\beta}) + \tfrac{K}{2}(\partial_{\beta}Q_{\alpha \beta})^2$, with \begin{equation}\label{eq:fed_bulk}
F(Q_{\alpha\beta}) = A_0\left(1-\frac{\gamma}{3}\right)\frac{Q_{\alpha \beta}^2}{2}-\frac{\gamma}{3}Q_{\alpha \beta}Q_{\beta \gamma}Q_{\gamma \alpha} + \frac{\gamma}{4}(Q_{\alpha \beta}^2)^2
\end{equation}
where Greek indices denote Cartesian coordinates and summation over repeated indices is implied. $A_0$ is a free energy scale, $\gamma$ is a temperature-like control parameter giving a order/disorder transition at $\gamma\sim 2.7$, and $K$ is an elastic constant. The anchoring at the particle surface is modeled by $f_s=W(Q_{\alpha\beta}-Q^{0}_{\alpha\beta})^2$, where $W$ is the anchoring strength and $Q^{0}_{\alpha\beta}$ is the preferred alignment of the nematic director at the particle surface.

The hydrodynamic equation for the evolution of $\mathbf{Q}$ is~\cite{berisedwards}:
$(\partial_t + u_{\nu}\partial_{\nu})Q_{\alpha \beta} - S_{\alpha\beta}= \Gamma H_{\alpha \beta}$,
where the first part describes the advection and $S_{\alpha\beta}$ describes the possible rotation/stretching of $\mathbf{Q}$ by the flow~\cite{berisedwards}. $\Gamma$ is the rotational diffusion constant and
the molecular field is
\begin{equation}
H_{\alpha \beta}= -{\delta 
  {\cal F} / \delta Q_{\alpha \beta}} + (\delta_{\alpha \beta}/3) {\mbox {\rm Tr}}({\delta {\cal F} / \delta Q_{\alpha \beta}}).
\end{equation}
The fluid velocity obeys $\partial_\alpha u_\alpha = 0$, and the Navier-Stokes equation, which is coupled to the LC via a stress tensor. We employ a 3D lattice Boltzmann algorithm to solve the equations of motion (for further details see {\it e.g.}~\cite{juho10, Lintuvuori10}).

{\it Simulation parameters:} We consider both pushers ($\beta < 0$) and pullers ($\beta > 0$). We fix the  $v_0=0.0015$, giving the particle velocity $u_0\equiv \tfrac{2}{3}v_0=10^{-3}$ in lattice units (LU), but vary the squirmer parameter in the range $\beta \in [-5,+5]$. We fix the  fluid viscosity $\eta = 0.1$ and the swimmer radius $R=4.0$ in LU (Fig. 1(a)). To model the nematic liquid crystal we use: $A_0 = 1.0$, $\gamma = 3.0$, $K = 0.01$, $\xi = 0.7$, $\Gamma = 0.3$ and a rotational viscosity $\gamma_1=\tfrac{2q^2}{\Gamma}=\tfrac{5}{3}$. The physics of our system is governed by the Reynolds (Re) and Ericksen (Er) numbers, which give the ratio of inertial and viscous forces, as well as the ratio of viscous and elastic force, respectively. Using the parameters above, we recover the following upper limits $\mathrm{Re} \equiv \tfrac{u_0R}{\eta}\approx 0.04$ and $\mathrm{Er}\equiv \tfrac{\gamma_1u_0R}{K}\approx 0.68$.
Simulations are carried in a rectangular simulation box $64\times 64\times 64$, with periodic boundary conditions.


\begin{figure}
\includegraphics[width=\columnwidth]{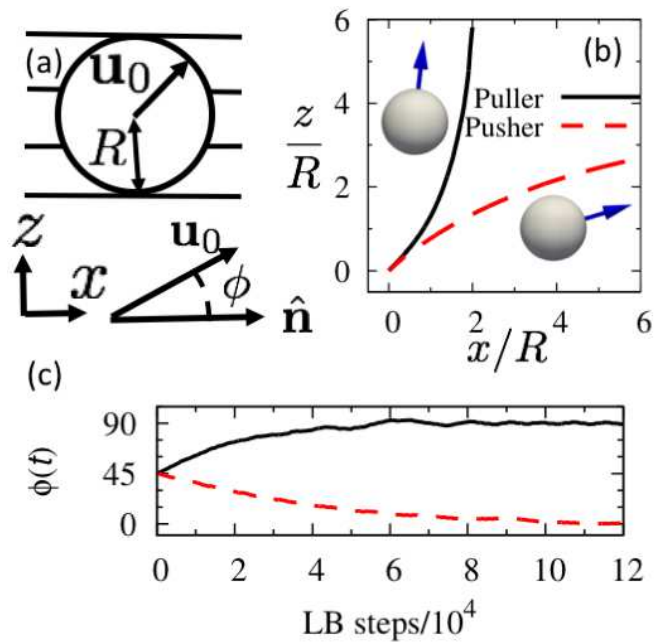}
\caption{(a) A cartoon showing the squirmer in a nematic liquid crystal, defining the angle $\phi$ used in the text, between the particle swimming direction $\mathbf{u}_0$ and the nematic director $\hat{\mathbf{n}}$. Examples of (b) the trajectory in $x-z$ plane and (c) the $\phi(t)$ observed in simulations of a puller ($\beta = +0.2$) and of a pusher ($\beta = -0.2$), with an initial orientation $\phi_0=45^{\circ}$.}
\label{Fig:AnglePM5}
\end{figure}

{\it Results:}
First we consider the case where the particle surface does not impose an alignment of the the nematic director ($W=0$). We place a single swimmer into a nematic liquid crystal with an initial angle $\phi_0=45^{\circ}$ between the squirmer orientation  and $\hat{\mathbf{n}}$ (Fig. 1(a)).
For a puller ($\beta = +0.2$), the hydrodynamically induced torque rotates the particle away from the nematic director leading to a curved trajectory towards a direction perpendicular to $\hat{\mathbf{n}}$ (Fig. 1(b) and (c) solid line). A $\beta = -0.2$ pusher instead starts to turn in the opposite direction, leading to swimming in the direction set by the nematic director (Fig. 1(b) dashed line), reaching a a steady state orientation $\phi\approx 0$ (Fig. 1(c) dashed line). (See also~\cite{Suppl} for additional movies of the puller and pusher.) 

The alignment of a pusher resembles the observation of bacterial swimmers in nematic LCs~\cite{Smalyukh08, Kumar13, Aranson14, Abbott14, Weibel15}, which are known to be rod-like pushers. However, for rod-like swimmers there exists an elastic energy penalty of re-alignment which depends on the orientation $\phi$ with respect to the $\hat{\mathbf n}$ and it is minimised when they align along $\hat{\mathbf n}$. Resulting elastic torque has been estimated as $T_{\mathrm{elastic}}\sim 4\pi K\phi L\ln(2L/R)\sim10^5\phi\mathrm{pN}\cdot\mathrm{nm}$~\cite{Smalyukh08}, which is considerably larger than that typically generated by the bacteria themselves $\sim 10^3\mathrm{pN}\cdot\mathrm{nm}$~\cite{Berg07, Smalyukh08}. Thus rod-like swimmers are expected to always align along $\hat{\mathbf n}$ independently of their swimming mechanism. On the contrary, for spherical swimmers $T_{\mathrm{elastic}} = 0$ thus any torque must arise solely from hydrodynamic interactions. 

To analyse the underlying physical mechanism, we discuss how the squirmer's flow field $\mathbf{v}(\mathbf{r})$ interacts with a liquid crystal in terms of the nematohydrodynamic equations~\cite{Suppl}. We study the torque exerted on the moving particle, 
\begin{equation}
\mathbf{T}=\oint\mathbf{r}\times\mathbf{\sigma}\cdot d\mathbf{S},
\end{equation}
where the integral runs over the particle surface having oriented surface element $d\mathbf{S}$.
A squirmer moving in an isotropic fluid with a viscosity $\eta_{\text{iso}}$ has a flow field $\mathbf{v}_{\text{iso}}(\mathbf{r})$. The viscous stress is defined as a linear function of velocity derivatives,
  $\mathbf{\sigma}_\text{iso}=\eta_\text{iso} \mathbf{\nabla}\mathbf{v}_\text{iso}$,
  and for a spherical particle one has $\mathbf{T}=0$. In a liquid crystal, the viscosity is an anisotropic fourth-rank tensor $\boldsymbol{\eta}$, and the stress is a rather intricate function of the strain $ \mathbf{\nabla}\mathbf{v}$ and the the order parameter $\mathbf{n}$. There is no analytical result for the squirmer velocity field $\mathbf{v}(\mathbf{r})$ in LC~\cite{Stark01}. 
We resort to a simple approximation that consists in evaluating the stress with the anisotropic viscosity $\boldsymbol{\eta}$ (given by Leslie coefficients $\alpha_i$ for nematic LC~\cite{deG93,Suppl}) but using the velocity field $\mathbf{v}_{\text{iso}}$~\cite{Mon12}, in the limit of small Reynolds and Ericksen numbers. 

From the velocity field of a moving squirmer $\mathbf{v}_{\text{iso}}$, we readily obtain the stress and the nemato-hydrodynamic torque exerted on the particle (for detailed calculation see supplementary material~\cite{Suppl}). The anisotropic part of the viscous stress is  dominated by $\mathbf{\sigma}-\mathbf{\sigma}_\text{iso}\propto\beta\boldsymbol{\eta}\hat{\mathbf{n}}(\hat{\mathbf{n}}\times\mathbf{\omega})$, where $\mathbf{\omega}=\mathbf{\nabla}\times\mathbf{v}_\text{iso}$ is the vorticity of the flow field and $\boldsymbol{\eta}$ the viscosity tensor. Inserting the known velocity field of a squirmer, we obtain the torque \cite{Suppl} 
\begin{equation}
  \mathbf{T}_N  = - 8\pi\beta\hat{\eta}v_0R^2  (\hat{\mathbf{n}}\cdot \hat{\mathbf{u}}) \hat{\mathbf{n}}\times \hat{\mathbf{u}},
  \label{eq:torque}
\end{equation}
where $\hat{\mathbf{u}}$ is the particle axis. The effective viscosity coefficient $\hat{\eta} = \tfrac{\alpha_1}{35} + \tfrac{\alpha_2 + \alpha_3}{2} + \tfrac{\alpha_5+\alpha_6}{20}$ is expressed in terms of the Leslie parameters $\alpha_i$ of a nematic liquid crystal~\cite{Suppl} and is dominated by the coefficients $\alpha_{2,3}$ related to the rotational viscosities, while $\alpha_{1,5,6}$ corresponds to shear viscosities~\cite{deG93,Suppl}.
When $\beta\hat{\eta}>0$, the torque aligns the particle axis on the order parameter. Throughout this paper we assume $\hat\eta<0$, which corresponds to measured values for common nematic LCs {\it e.g.} 5CB and MBBA and to the simulations~\cite{Suppl}. Then Eq. (\ref{eq:torque}) predicts that the stable orientation of pullers ($\beta>0$)  is perpendicular to the nematic order, whereas pushers ($\beta<0$) move in the parallel direction. To test this predictions, we carried out simulations for a $\beta = +5$ puller  and a $\beta = -5$ pusher, and initialised the system close to the unstable orientation. Fig. 2a shows for the evolution of $\phi(t)$ an S-shaped trajectory, towards the stable positions given by (eq.~\ref{eq:torque}). 

To determine the angular velocity $\Omega$, we match the torque $T_N$ with the friction induced by the particle's rotation, $T_N - 8\pi\hat\eta_\Omega R^3 \Omega=0$, with the viscosity $\hat\eta_\Omega$ of rotational Stokes drag \cite{Suppl}. Noting that the scalar and vector products in (\ref{eq:torque}) result in a factor $\cos\phi\sin\phi=\frac{1}{2}\sin(2\phi)$, we find
\begin{equation}
  \Omega= -\frac{1}{2}\beta\sin(2\phi)\frac{\hat\eta}{\hat\eta_\Omega}\frac{v_0}{R}.
  \label{eq:rotvel}
\end{equation}
The scale is given by $v_0/R$, and $\Omega$ is proportional to the squirmer parameter $\beta$ and varies with the angle $\phi$.  In Figure 2(b), this is compared with the numerical derivative $\Omega(\phi)=d\phi/dt$ from the simulation data for the $\beta=+5$ puller. The data shows very good agreement between theory and numerics. Starting from the initial position $\phi=0$, the simulated velocity increases linearly with $\phi$, then reaches a maximum at $\phi\approx 45^{\circ}$ and finally slows down when approaching the stable orientation $\phi=90^{\circ}$.

\begin{figure}
\includegraphics[width=\columnwidth]{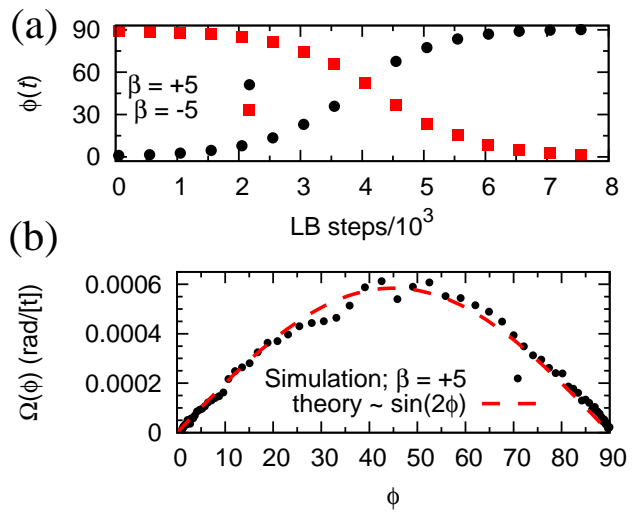}
\caption{(a) $\phi(t)$ exhibits an S-shaped evolution, with a stable configuration $\phi\approx 90^{\circ}$ ($\phi\approx 0^{\circ}$) for a puller (pusher). (b) The rotational velocity, $\Omega(\phi)$ is symmetric around $\phi = 45^{\circ}$ and vanishes for $\phi\rightarrow 0$ and $90^{\circ}$, in agreement with theoretical arguments (see text for details).}
\label{Fig:AnglePM5}
\end{figure}

Modifying the squirmer parameter $\beta$ keeps the swimming speed constant, but changes the power of the squirmer flow field~\cite{brown15}. This far we have established that the sign of $\beta$ defines the stable swimming direction with respect to the nematic axis. To understand how the magnitude of the the hydrodynamically induced torques depend on the power of the squirmer flow field, we initialise the simulations with $\phi_0 = 45^{\circ}$, and systematically vary $\beta$ between -5 and +5. We evaluate the $\Omega(\beta)$ from a linear fit to early times on $\phi(t)$ data (see {\it e.g.} early times in Fig. 1(c)). The $\Omega(\beta)$ from simulations shows a linear dependence for all the values of $\beta$ considered (Fig. 3) and indeed the theory predicts $\Omega(\beta)\sim \beta$ for a fixed $\phi$ (see {\it e.g.} eq.~(\ref{eq:rotvel})). 

\begin{figure}
\includegraphics[width=\columnwidth]{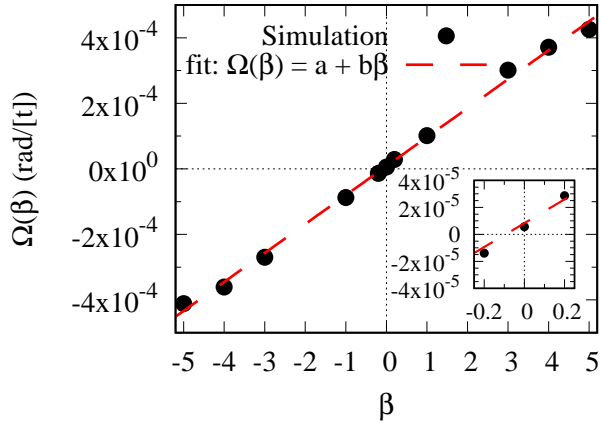}
\caption{The rotational velocity $\Omega(\beta)$ shows a linear dependence of the squirming parameter $\beta$ for a fixed $\phi$. For $\beta = 0$, $\Omega$ takes a value $a\approx 8\times10^{-6}\tfrac{\text{rad}}{[t]}$ (see inset and text for details). 
  }
\label{Fig:RotVel}
\end{figure}

  Interestingly our numerical simulation results show that the re-orientation dynamics for pullers  is slightly more rapid than for pushers (see Fig. 1c and inset in Fig. 3 for $\beta =\pm 0.2$). 
  Also the angular velocity shown in Fig. 3, does not vanish at $\beta = 0$ (inset Fig. 3) but in a steady state a neutral squirmer swims perpendicular to $\hat{\mathbf{n}}$ (See the supplement~\cite{Suppl} for $\phi(t)$ for $\beta = 0$). This behaviour is not captured by our analytics.
  In our theoretical treatment we replace $\mathbf{v}(\mathbf{r})$ with the velocity field calculated in an isotropic liquid $\mathbf{v}_\text{iso}(\mathbf{r})$.
  The analytical results agree remarkably well with the (more precise) numerical simulations, concerning the dependencies of $\Omega$ on the squirmer parameter $\beta$ and the orientational angle $\phi$ (see {\it e.g.} Fig. 2(b) and Fig. 3). 
  This is in line with a previous study of anisotropic diffusion of colloids, where this approximation was shown to compare favorably with numerically exact results \cite{Mon12}.
  The reorientation of $\beta =0$ swimmer could probably be reproduced when using the exact velocity field $\mathbf{v}(r)$, which depends itself on the viscosity anisotropy $\boldsymbol{\eta}$.

\begin{figure}
\includegraphics[width=\columnwidth]{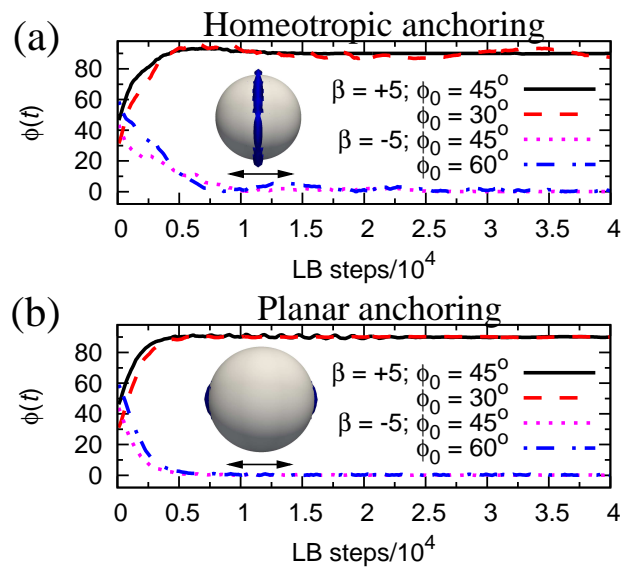}
\caption{The angle $\phi(t)$ for particles with (a) homeotropic and (b) planar anchoring of the nematic director at the particle surface ($WR/K\approx 4$) for both a puller ($\beta = +5$) and a pusher ($\beta = -5$). The insets show the defect structure around a passive particle: (a) Saturn ring defect for a homeotropic anchoring and (b) two boojums for a planar anchoring at the particle surface. (The arrow denotes the orientation of the far-field nematic director.)}
\label{Fig:AnglePMAnchoring}
\end{figure}

In all the examples above, we have considered a case where there is no anchoring at the nematic director at the surface of the colloidal particle ($W=0$).
Typically in experiments the particle surface interacts with the nematic director ($W>0$).
The case of homeotropic anchoring can lead to the formation of a Saturn ring defect near the particle surface (see e.g. inset in Fig. 4(a)). In the case of degenerate planar anchoring, two boojums are observed at both poles of the particle (inset of Fig. 4(b)).  We still observe the reorientation of the squirmers when a reasonably strong surface anchoring is included ($WR/K = 4$), 
as shown in Fig. 4 for $\beta = + 5$ puller and $\beta = - 5$ pusher. This provides further evidence that the re-orientation is due to the hydrodynamic coupling between the squirmer flow field and the anisotropic viscosities of the LC, as opposed to short range elastic interactions.

Our main finding is that nematic liquid crystal exerts a torque on a spherical microswimmer. This should be easily observable in experiments. Using typical values for the LC viscosities~\cite{deG93,Suppl}, and for microswimmers ($R\sim 1\mu$m and $v_0\sim 1\ldots10\mu$m/s), we can estimate the magnitude of the torque (eq.~\ref{eq:torque})  $T\sim 4\beta\times(10^2\ldots 10^{3})\mathrm{pN}\cdot\mathrm{nm}$, and $\Omega\sim \beta \tfrac{\mathrm{rad}}{s}$, which is comparable to the recently observed rotation induced by a laser intensity gradient on a thermally powered Janus particle~\cite{Loz16, Bic14}. Further, the reorientation rate $\Omega$ is much faster than typical rotational diffusion. These, combined with the observation that the steady state behaviour is retained for $\tfrac{WR}{K}>0$, suggest that our prediction should be testable in the laboratory, for example by dispersing artificial swimmers {\it e.g.} ~\cite{brown14, brown15, ebbens12b, sabass10, ebbens12, ebbens14, wang15, das15, Loz16} into standard nematic liquid crystals {\it e.g.} 5CB or MBBA.


{\it Conclusions:}
We have presented a combined simulation and analytical calculation study of the steady state swimming of a spherical squirmer in a nematic liquid crystal.
In a steady state a pusher will swim along the nematic director while a puller will be moving perpendicular to the direction set by the far-field $\hat{\mathbf{n}}$. We show via analytic calculations that the re-orientation of the swimmers arises from the hydrodynamic coupling between the squirmer flow field and the anisotropicity of the liquid crystalline viscosities. For a passive spherical colloidal particle moving slowly in a thermotropic nematic LC 
a ratio of viscosities parallel ($||$) and perpendicular ($\perp$) to $\hat{\mathbf{n}}$ has been observed $\tfrac{\eta_{\perp}}{\eta_{||}}\approx 2$ experimentally~\cite{Loudet04, Tiffany}, and by both theoretical calculations and simulations~\cite{Stark01, Lintuvuori10}. 
Our calculations show that the anisotropy of the liquid crystal viscosities~\cite{miesowicz} gives a rise to a hydrodynamic torque on the squirmer, leading to the observed steady state behaviour.
Finally, the steady state behaviour persists even when a strong anchoring of the LC director at the particle surface is included, rendering it directly experimentally relevant. The predictions should be valid for spherical microswimmers. 

A good candidate for an experimental realisation of predictions would be to consider lyotropic nematic liquid crystal~\cite{Mon12}, for both artificial or bacterial swimmers. Here, recent experiments of a diffusion of colloidal particles showed a viscosity ratio $\eta_{\perp}/\eta_{||}\sim 4$~\cite{Mon12}, which is larger anisotropy than considered here. Using thermotropic (oil-based) LCs, would require particles capable swimming in oil. The predictions presented here could also be valid for a wider class of materials which exhibit anisotropic viscosities, {\it e.g.} lyotropic lamellar phases could an interesting host material for future studies.
Our results suggest that anisotropic materials could offer an exciting, yet easy-to-use, platform to guide microswimmers in the bulk. This could allow for example directed transport, or sorting of swimmers based on their hydrodynamic signature, by simply dispersing them into an environment with a broken symmetry ({\it e.g} nematic liquid crystal.)

{\it Acknowledgments:} JSL acknowledges funding from IdEx Bordeaux. AW acknowledges support by Agence Nationale de la Recherche through contract ANR-13-IS04-0003.



\onecolumngrid
\newpage

\makeatletter 
\def\tagform@#1{\maketag@@@{(S\ignorespaces#1\unskip\@@italiccorr)}}
\makeatother

\makeatletter \renewcommand{\fnum@figure}
{\figurename~S\thefigure}
\makeatother

\setcounter{equation}{0}
\setcounter{figure}{0}

\begin{center}
  {\Large \bf Supplementary information for squirmers in nematic liquid crystal}
\end{center}

\medskip

\section{Calculation of the Leslie viscosities}

For the lattice Boltzmann model presented in the main text, the Leslie viscosities of a passive nematic liquid crystals are defined in the terms of, the rotation diffusion constant $~\Gamma$, the $\tfrac{3}{2}$ of the largest eigenvalue of the order parameter tensor $\mathbf{Q}$, $q$, flow alignment parameter $\xi$ and isotropic viscosity $\eta$. Using the values from the simulations in the main text $\Gamma = 0.3$, $q=\tfrac{1}{2}$, $\xi = 0.7$ and $\eta = 0.1$, we recover following Leslie viscosities ~\cite{deG93, davide07} in simulation units:
\begin{align}
  \alpha_1 &= -\frac{2}{3\Gamma}q^2(3 + 4q - 4q^2)\xi^2 &&\approx -1.09, \\
  \alpha_2 &= \frac{1}{\Gamma}\left(-\frac{1}{3}q(2+q)\xi - q^2\right) &&\approx -1.81, \\
  \alpha_3 &= \frac{1}{\Gamma}\left(-\frac{1}{3}q(2+q)\xi + q^2\right) &&\approx -0.14, \\
  \alpha_4 &= \frac{4}{9\Gamma}(1-q)^2\xi^2 + \eta &&\approx 0.28, \\
  \alpha_5 &= \frac{1}{3\Gamma}\left[q(4-q)\xi^2 + q(2+q)\xi\right] &&\approx 1.93, \\
  \alpha_6 &= \frac{1}{3\Gamma}\left[q(4-q)\xi^2 - q(2+q)\xi\right] &&\approx -0.02.
\label{alpha model}
\end{align}
These give $\tfrac{\alpha_2 + \alpha_3}{2}\left(\approx -0.98\right) < 0$ as well as
$\left|\tfrac{\alpha_2 + \alpha_3}{2}\right| >> \left|\tfrac{\alpha_5 + \alpha_6}{20}\right|\left(\approx 0.098\right)$
and
$\left|\tfrac{\alpha_2 + \alpha_3}{2}\right| >> \left|\tfrac{\alpha_1}{35}\right|\left(\approx 0.03\right)$,
as required in the main text.

For comparison, we give measured values for liquid crystals 5CB (MBBA) of the coefficients $\alpha_2+\alpha_3=-107$ $(-79)$  mPa.s,  
$\alpha_5+\alpha_6=107$ $(81)$  mPa.s,   $\alpha_1=-11$ $(7)$  mPa.s \cite{deG93}. The isotropic term takes the value  $\alpha_4=75$ $(83)$  mPa.s.

\section{Neutral squirmer}
\begin{figure}[h!]
\includegraphics[width=0.5\columnwidth]{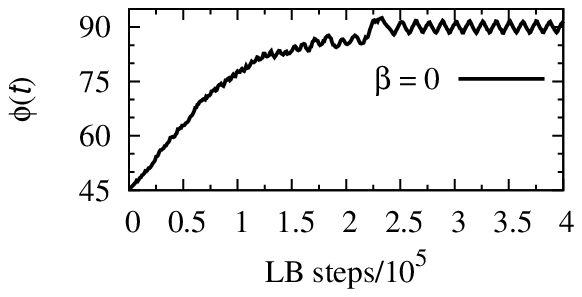}
\caption{The time evolution of the angle $\phi(t)$ between the squirmer orientation and nematic director $\hat{\mathbf{n}}$, shows that neutral squirmer ($\beta = 0$) orients perpendicular to the far-field nematic director.} 
\label{Fig:AngleB0.0}
\end{figure}

\section{Nematohydrodynamics}

The action of a moving fluid on a dispersed particle is given by well-known
relations for the force
\begin{equation}
\mathbf{F}=\oint\mathbf{\sigma}\cdot d\mathbf{S} \label{32}%
\end{equation}
and the torque
\begin{equation}
\mathbf{T}=\oint\mathbf{r}\times\mathbf{\sigma}\cdot d\mathbf{S}, 
\label{34}%
\end{equation}
where the integral is over the particle surface with position vector
$\mathbf{r}$\ and the oriented surface element $d\mathbf{S}=-dS\mathbf{\hat{r}}$.

The  stress tensor $\sigma_{ij}=\sigma_{ij}^{\prime}-P\delta_{ij}$ consists of a viscous part 
$\sigma_{ij}^{\prime}$ and the pressure $P$. 
In the framework of the stationary Leslie-Ericksen equations the viscous stress reads as \cite {deG93}
\begin{equation}
\mathbf{\sigma}^{\prime}  =\alpha_{1}\mathbf{nn\cdot E \cdot nn} + \alpha_{4} \mathbf{E}
  +\frac{\alpha_{5}+\alpha_{6}}{2}\left(  \mathbf{nn\cdot E} + \mathbf{E \cdot nn}\right) 
  +\frac{\alpha_{3}+\alpha_{2}}{2}\left( \mathbf{n N} + \mathbf{Nn}\right),
\label{10}
\end{equation}
with the strain tensor 
$$\mathbf{E}=\frac{1}{2} \left(\mathbf{\nabla  v} + (\mathbf{\nabla  v})^\dagger\right),$$  
the vorticity 
$$\mathbf{\omega}= \mathbf{\nabla } \times \mathbf{v},$$  
and the rate of change  of the nematic order parameter
$$ \mathbf{N} = [ (\mathbf{v} - \mathbf{u})\cdot\mathbf{\nabla}] \hat{\mathbf{n}} + \frac{1}{2}\hat{\mathbf{n}}\times\mathbf{\omega}.$$

\section{Approximations}

With (\ref{10}) it is not possible to solve Stokes' equation ${\bf\nabla}\cdot{\bf\sigma}=0$. 
In order to obtain a problem that is tractable in a simple analytical
approach, we resort to the following approximations. First, we neglect the
deformation of the nematic order due to the interaction with the particle's
surface and with its velocity field. This corresponds to case of no surface anchoring ($W=0$) and the limit of a small Ericksen number, $\text{Er}\to 0$. In other words we assume that the order
parameter is constant in space and time,
\begin{equation}
\hat{\mathbf{n}}=\text{const.}%
\end{equation}
Then the rate of change of the order parameter simplifies as
$$\mathbf{N}=\frac{1}{2}\hat{\mathbf{n}}\times\mathbf{\omega}.$$

Second, we calculate the viscous stress (\ref{10}) with the velocity field of an 
active particle in an isotropic fluid $\mathbf{v}_{\text{iso}}(\mathbf{r})$. Writing the lowest terms of the well-known series 
as gradient and and rotational fields, we have
\begin{equation}
\mathbf{v}_\text{iso}    = \mathbf{\nabla}\Phi + \mathbf{\nabla} \times \mathbf{A},
\label{36}
\end{equation}
with the scalar
\begin{equation}
\Phi   = - v_0 \left(\frac{R^3\hat{\mathbf{r}}\cdot\hat{\mathbf{u}}}{3 r^2} + \beta  \frac{R^4P_2(\hat{\mathbf{r}}\cdot\hat{\mathbf{u}})}{3 r^3} \right) 
         =  - v_0 \left(\frac{R^3 \cos\theta}{3 r^2} + \beta  \frac{R^4P_2( \cos\theta)}{3 r^3} \right) 
\end{equation}
and the vector field 
\begin{equation}
\mathbf{A} = \beta v_0\hat{\mathbf{u}}\times\hat{\mathbf{r}}\frac{R^2(\hat{\mathbf{r}}\cdot\hat{\mathbf{u}})}{2 r}
                   =   \beta v_0 \frac{R^2 \sin\theta \cos\theta}{2 r} \mathbf{e}_\varphi ,
\end{equation}
where we have defined the unit vectors $\hat{\mathbf{r}}=\mathbf{r}/r$ and $\hat{\mathbf{u}}=\mathbf{u}/u$. The polar angle $\theta$ and the azimuthal unit vector $\mathbf{e}_\varphi$ are defined with respect to the particle axis $\hat{\mathbf{u}}$.  

The above relations give the flow field for an isotropic fluid, which is calculated with the
viscosity parameter $\alpha_{4}$ only. Thus the viscosity
anisotropy appears only in the prefactors in (\ref{10}), and we neglect that
the velocity field itself depends on the anisotropic viscosity parameters
$\alpha_{1},\alpha_{2},\alpha_{3},\alpha_{5},\alpha_{6}$.

Measured values for the anisotropy parameters are not small as compared to the isotropic viscosity 
$\alpha_4$ \cite{deG93}. In  previous work, this approximation was used for evaluating the anisotropy 
of Brownian motion of spherical particles in zero anchoring conditions. Comparison with numerically 
exact results \cite{Sta01} revealed an error of less than ten percent \cite{Mon12}, suggesting this 
approximation to be rather robust.

Spelling out the about derivatives one readily obtains the explicit form of $\mathbf{v}_\text{iso}$. From the expression at the surface $r=R$, one finds the particle velocity $u_0=-\frac{2}{3}v_0$ and the slip velocity 
\begin{equation}
 v_s = v_0 \sin\theta \left(1 + \beta \cos\theta \right) ,
\end{equation}
with the squirmer parameter $\beta$.
Note that $\mathbf{v}_\text{iso}$ is given in the laboratory frame and $v_{s}$ in the particle-fixed frame; both are related through $v_s=(\mathbf{v}_\text{iso}-\mathbf{u})\cdot(1-\hat{\mathbf{r}}\hat{\mathbf{r}})$. 

\section{Torque exerted on the particle}

Calculating the strain tensor from (\ref{36}) and inserting the symmetrized
stress tensor (\ref{10}), we evaluate the force and torque exerted on the
particle. Not surprisingly, (\ref{32}) vanishes, $\mathbf{F}=0$, since there
is no external potential.

The torque, on the contrary, takes a finite value. Inserting the velocity field in  (\ref{10}) and 
performing the integral in (\ref{34}), one can evaluate the torque. For $\hat{\mathbf{n}}$ along $x$-axis and the particle moving in the $x-z$-plane, we find 
\begin{equation}
\mathbf{T}_N = -4\pi \beta\sin(2\phi)\hat{\eta} v_0 R^{2}\mathbf{\hat{y}}
                         =- 8\pi\beta\hat{\eta}v_0R^2  (\hat{\mathbf{n}}\cdot \hat{\mathbf{u}}) \hat{\mathbf{n}}\times \hat{\mathbf{u}} ,
\label{42}
\end{equation}
with $\phi$  denoting the angle between the particle axis and the order parameter, 
$\cos\phi=\hat{\mathbf{u}}\cdot\hat{\mathbf{n}}$, and the unit vector $\mathbf{\hat{y}}$ 
which is perpendicular to both  $\mathbf{u}$ and  $\hat{\mathbf{n}}$. The viscosity coefficient
reads 
\begin{equation}
\hat{\eta}=\frac{\alpha_{1}}{35}+\frac{\alpha_{2}+\alpha_{3}}{2}%
+\frac{\alpha_{5}+\alpha_{6}}{20}.\label{44}%
\end{equation}
Not surprisingly, the isotropic viscosity $\alpha_{4}$ does not contribute;
$\hat{\eta}$\ is determined by the anisotropy parameters $\alpha_{1}%
,\alpha_{2},\alpha_{3},\alpha_{5},\alpha_{6}$.

The torque is perpendicular on the order parameter $\hat{\mathbf{n}}$ and the
particle velocity $\mathbf{u}$; it vanishes both for parallel and
perpendicular orientations, $\phi=0$ and $\phi=\frac{\pi}{2}$.

With the viscosity parameters of commonly used LC, such as 5CB and MBBA, one
finds that $\hat{\eta}$ is determined by the coefficient $\alpha_{2}%
+\alpha_{3}$. In other words, the interaction of an active particle and the
nematic order is dominated by the last term of the viscous stress (\ref{10}),
which in a case of constant $\hat{\mathbf{n}}$ accounts for the vorticity of the
fluid velocity field.

For the mentioned systems $\alpha_{2}+\alpha_{3}$ is negative. In view of
(\ref{42}) this means that for $\beta>0$ the particle axis $\mathbf{\hat{u}}$
is turned away from the order parameter, and that the stable orientation is
perpendicular to $\hat{\mathbf{n}}$. On the other hand, for a negative squirmer
parameter $\beta<0$, we expect that the active particle aligns on the nematic
order parameter.\ 

\section{Angular velocity of the squirmer particle}

The torque $\mathbf{T} $ exerted by the nematic order results in a angular velocity $\Omega$ of the squirmer particle, 
which in turn induces an additional flow field
\begin{equation}
\mathbf{v}_\Omega = \frac{  R^3}{r^3} \mathbf{\Omega}\times \mathbf{r} = \frac{\Omega  R^3}{r^2} \sin\theta \mathbf{e}_\varphi.
\label{51}
\end{equation}
The angular velocity adjusts such that the corresponding viscous torque exactly cancels that exerted by the 
nematic order, $\mathbf{T}_\Omega+ \mathbf{T}=0$. Here we use this relation in order to determine $\Omega$.

Calculating the strain tensor  $\mathbf{E}_\Omega$ and the vorticity vector $\mathbf{\omega}_\Omega$  we obtain  
the viscous stress (\ref{10}). Upon performing the surface integral we find the viscous torque 
\begin{equation}
\mathbf{T}_\Omega= - 8\pi\hat{\eta}_\Omega \Omega R^3 \mathbf{e}_y,
\label{52}
\end{equation}
with the effective viscosity 
\begin{equation}
\hat{\eta}_\Omega= \frac{\alpha_4}{2} + \frac{\alpha_2+\alpha_3}{20} + \frac{9(\alpha_5+\alpha_6)}{40} + \frac{\alpha_1}{10}.
\end{equation}
Identifying the elastic and viscous torques, we find the angular velocity of the squirmer
\begin{equation}
\Omega = -\beta \frac{\sin(2\phi)}{2} \frac{\hat\alpha}{\hat\alpha_\Omega} \frac{v_0}{R} \equiv -\sin(2\phi)\Omega_0. 
\label{52}
\end{equation}
Taking the ratio of the viscosity parameters equal to unity, $\hat\alpha\sim \hat\alpha_\Omega$, we find $\Omega_0=\beta v_0/2R$.

Here a remark on the validity of our approximative evaluation of viscous stresses is in order: Since the viscous torque is necessarily opposite 
to the driving velocity $\Omega$, the viscosity parameter needs to be positive; in other words, the anisotropy parameters should be small as 
compared to the isotropic viscosity $\alpha_4$. This condition is not fulfilled by the model parameters (\ref{alpha model}) nor by the values 
measured for common liquid crystals. As a consequence, the numerical values of the coefficients $\hat{\eta}$ and $\hat{\eta}_\Omega$ 
are probably subject to large uncertainties.

\end{document}